\documentstyle[11pt,psfig,rotate,aaspp4]{article}

\slugcomment{\it(To appear in The Astronomical Journal, September 1998)}

\begin{document}

\title{An Optical Multicolor System For Measuring \\
	Galaxy Redshifts And Spectral Types}

\author{Charles T. Liu\altaffilmark{1} and Richard F. Green}
\affil{National Optical Astronomy Observatories, Box 26732, Tucson, AZ 85726}
%

\altaffiltext{1}{Current address: Department of Astronomy, Columbia University, 550 W 120 St, New York, NY 10027}

\begin{abstract}

A method of obtaining approximate redshifts and spectral types
of galaxies using a photometric system of six broad-bandpass filters 
is developed.  The technique utilizes a smallest maximum difference
approach rather than a least-squares approach, and does not consider
a galaxy's apparent magnitude in the determination of its redshift.
In an evalution of its accuracy
using two distinct galaxy samples, the photometric redshifts 
are found to have a root mean square deviation of
$\pm$0.05 from spectroscopically determined redshifts.
Possible systematic errors
of the method are investigated, including the effects of post-starburst
(``E+A'') galaxies and attempts to measure redshifts with incomplete
color information.  Applications of the technique are discussed.

\end{abstract}


\section{Introduction}

The traditional
method of obtaining redshifts through spectroscopy, though accurate, is
by far the most time-consuming task in any observational study of faint
extragalactic objects.  This is especially true for galaxies, most of 
which do not have spectral features so dominant as to lend themselves to
quick, unambiguous redshift measurements.  The need to obtain large numbers
of redshifts for large samples of galaxies, however, has never been greater.
Many have therefore attempted to design and refine 
methods to obtain accurate redshifts of galaxies with very low
spectral resolution (see, e.g., Koo 1985; Loh \& Spillar 1986;
Connolly et al. 1995; Sawicki, Lin \& Yee 1997; Brunner et al. 1997).

We present here a photometric redshift technique we have developed, using 
template galaxy spectral energy distributions (SEDs) 
and a system of six optical 
and near-infrared broad-band filters.  The technique relies on a smallest
maximum difference approach rather than a least-squares method,
and is optimized for the
analysis of galaxy data obtained from the deep multicolor survey of Hall
et al. (1996a, hereafter HOGPW).  The 
results of this work are being applied to the 
data in a companion paper (Liu et al. 1998).

\section{Model Galaxies and Colors}

The six-band filter system of HOGPW
consists of standard Johnson U, B and
V filters, and non-standard R and I filters (called R, I75, and I86) with 
approximate effective wavelengths of 6615\AA , 7425\AA , and 8586\AA\
respectively, and FWHM of $\sim$1000\AA\ each.  Together, the system covers
the wavelength range 3000-9000\AA\ in approximately 1000\AA\ intervals.
The detector response included in the model colors was also that of the
one used in the HOGPW survey, {\it i.e.} a Textronix 2048 $\times$ 2048 
CCD.  The transmission functions of the filters, and the quantum efficiency
curve of the CCD, are presented in Figure 1.

Five representative galaxy templates are
used in this photometric redshift scheme.  
Four of the models are taken from the
integrated spectrophotometry of Kennicutt (1992) and Coleman, Wu \&
Weedman (1980), for spectra 
representative of E/S0, Sbc, Scd and Irr (starburst) galaxies.  
The higher-resolution spectra from Kennicutt (1992) were spliced
together with the Coleman et al. data --- NGC 5248 with the Sbc galaxy, 
NGC 6181 with the Scd galaxy, and NGC 4449 with the Irr galaxy --- to create 
continuous spectral energy distributions. 
The fifth SED is a composite spectrum of Sa and Sab galaxies from
Kinney et al. (1996).  Our selection of these 
templates was motivated by our desire to use data of real galaxies,  
to avoid the uncertainties involved with spectra created artificially
by stellar population models.  This is also why we have not used
interpolated galaxy spectra to create a finer mesh of SEDs.

The SEDs are presented in Figure 2.  We will
be using the term ``galaxy spectral type'' often in this 
paper; this simply means we are referring to galaxies with the same
SEDs, and hence the same implied star formation rates, as those of the 
representative galaxies mentioned above.

We computed the colors of our template galaxies by first placing them
onto a standard photometric system, to derive colors that could be used to
test data on that system.  Thus, we used the spectrophotometry of
Hamuy et al. (1994), who have produced SEDs of a number of Southern
Hemisphere standard stars across the UBVRI wavelength range. We convolved
the SEDs of 14 of these stars with our filter and CCD response functions,
to produce an ``instrumental flux'' through each filter, $I_{f}$:

\begin{equation}
I_f = \displaystyle \int T_f(\lambda) D(\lambda) F(\lambda) d\lambda
\end{equation}

\noindent
where $T_f(\lambda)$ is the transmission of filter $f$, $D(\lambda)$ is
the quantum efficiency of the CCD, and $F(\lambda)$ is the flux of the star.
In other words, we were measuring a net flux from each star as if
it had been observed through a telescope with a flat wavelength 
throughput of unity, with the exception of the filter and detector,
and with no atmospheric extinction.

For the UBVR filters, we then used the photometric calibration 
of those stars by Landolt (1992) to run the PHOTCAL 
package in IRAF.
Using this procedure, we obtained the photometric zero points and
color terms with which we could transform any data
simulated by our models onto the same standard UBVR system as HOGPW.
The I75 and I86 filters are calibrated onto an absolute scale by HOGPW
using the standard star Wolf 1346.  To determine the
transformation coefficients for these two filters, we processed the
SED of Wolf 1346 through our procedure, and compared
the instrumental flux with the absolute calibrated magnitudes defined in
HOGPW to obtain the zero points.  (Color terms were not used in the
calibration of the I75 and I86 filters.)

The final derived parameters of the transformation from an ``instrumental''
magnitude to the HOGPW UBVRI75I86 photometric system, calibrated onto an
absolute scale with Wolf 1346, is as follows:

\centerline{ U$_{inst}$ = U $-$ 17.79 $-$ 0.042(U$-$B) }
\centerline{ B$_{inst}$ = B $-$ 19.13 $-$ 0.047(B$-$V) }
\centerline{ V$_{inst}$ = V $-$ 20.02 $-$ 0.042(V$-$R) }
\centerline{ R$_{inst}$ = R $-$ 19.74 $-$ 0.042(V$-$R) }
\centerline{ I75$_{inst}$ = I75 $-$ 19.58 }
\centerline{ I86$_{inst}$ = I86 $-$ 19.26 }

The formal errors for the transformation, as computed by PHOTCAL, are
less than $\pm$0.006 in magnitude and $\pm$0.013 in the color term for each
individual bandpass.  Each model magnitude thus has an error of at most a few 
percent.  The same is true for any color we choose to compute from these
magnitudes.

The expected observed colors as the galaxies increase with
redshift from $z$=0 to 1.0, in steps of $\Delta z$=0.025,  were
produced by applying K-corrections to the
rest frame colors.  The corrections were computed with direct numerical 
integration in the standard way:

\begin{equation}
K_f = 2.5\ log\ (1+z) + 2.5\ log\ \frac{\int F(\lambda)T_f(\lambda)D(\lambda)d(\lambda)}
{\int F[\lambda/(1+z)]T_f(\lambda)D(\lambda)d(\lambda)}
\end{equation}

\noindent
We present our computed model galaxy colors for the five galaxy spectral 
types in Tables 1-5.  To check our models, we compared them with the galaxy
color indices from the large photometric surveys of Poulain \& Nieto (1994),
Buta \& Williams (1995), and de Jong (1995).  The template colors agree
very well with median and mean color indices for the corresponding
Hubble types, with at most $\pm$0.05 magnitude variations.

\section{The Photometric Redshift Method}

The photometric redshift method should, in principle, output both a redshift
and a galaxy spectral type for each set of input from a galaxy.
As demonstrated by Koo (1985), Connolly et al. (1995) and others,
a combination
of sufficient wavelength coverage (i.e. near-UV to near-IR) and data from
at least four filters, such as U/B$_J$/R$_F$/I$_N$, is sufficient to obtain
redshifts with roughly $z \pm 0.05$ accuracy.  Thus the filter set we
use here, which contains six filters and roughly the same wavelength coverage,
is theoretically more than sufficient for the task of finding 
photometric redshifts.  Additionally, the division of the Johnson
I-band into two narrower bandpasses (I75 and I86) gives us additional leverage
at higher redshifts, as prominent spectral features (especially
the 4000 \AA\ break) are redshifted into that wavelength range.

As a first step, we need to know how well the various types of 
galaxies are separated in our multi-dimensional color space.  In other words,
how unique are the UBVRI75I86 colors of a galaxy with a given redshift and
spectral type?  A straightforward analysis shows that, in the 
(U-B)/(B-V)/(V-R)/(R-I75)/(I75-I86) color space, every galaxy 
redshift-spectral type pair on the previously stated binning interval
is separated from every other such pair by
at least 0.1 magnitude in at least one color; and the distinctions across
broader wavelength ranges (e.g. B-R or R-I86) are even more pronounced.

The separation of SEDs in the multicolor space is most easily illustrated
by tracing where the redshift-spectral type pairs lie on color-color diagrams.
Figure 3 shows two color-color cuts in the multicolor space of the 
filter system,
and the locations of the galaxy redshift-spectral type loci on them.  These
tracks can be thought of as color tracks for non-evolving 
galaxies as a function of redshift.  In the (U-B)/(B-R) plane, galaxies
with redshifts $z\leq$0.7 are well separated.  Similarly, 
in the (B-V)/(R-I86) plane, galaxies with $z\geq$0.4 are well separated.  A
representative shift in the colors that would be created by extinction is shown
with a reddening vector for E(B-V)=0.1.

These diagrams are also useful because they show the situations where the
photometric redshift method is most and least effective.  Early-type (E/S0
and Sab) galaxies are very easily separated from the other types, 
and move significantly
in color space as a function of redshift; clearly, these galaxies are the
ones most easily identified with this technique -- a well-known fact
that has been used by many authors (e.g. Im et al. 1996).
Higher-redshift blue galaxies (especially
irregulars), on the other hand, are somewhat problematic; increases in 
redshift around 0.4$\lesssim z\lesssim$0.6, and near $z=$1, only slightly 
change the observed colors of these galaxy types.  They are still 
unlikely to be confused with different galaxy spectral types; the 
intrinsic errors of their redshift determinations, however, will be larger.
Some scatter can also come from variations in emission-line strength, 
particularly the H$\alpha +$ N[II] complex for starburst galaxies; but
the overall effect is unlikely to be more than 0.05-0.1 magnitude in any
given color, which is less than the difference seen between almost all
of the different redshift-type pairs on our grid of models.

\section{Comparison with Spectroscopy}

The acid test of photometric redshifts is to compare them with
spectroscopic redshifts.  We used two distinct samples of galaxies,
obtained independently and in different ways, and 
examined how well the photometric measurements
match the spectroscopy separately and as a whole.
This was done to see if our system yields the
same results with different datasets obtained with this filter set,
which is what we expect.

\subsection{Cluster Galaxy Data}

UBVRI75I86 photometry were obtained for four rich clusters of galaxies:
Abell 963 ($z=$0.20), CL1358+6245 ($z=$0.32), CL3C295 ($z=$0.46), and
CL1601+4253 ($z=$0.54).  These clusters have numerous spectroscopic 
redshift measurements of cluster members of all spectral types,
as well as some foreground and
background galaxies (cf. Lavery \& Henry 1988; Fabricant, McClintock \&
Bautz 1991; Dressler \& Gunn 1992), and a large range of redshifts, making
them very desirable for testing the photometric redshift technique.

CCD imaging observations were made with the Steward Observatory
2.3-m telescope on Kitt Peak.  The central
2$\arcmin \times$ 3$\arcmin$ of each cluster were imaged in UBV
with a thinned, blue-sensitive 800$\times$1200 CCD, and the central 
5$\arcmin \times$ 5$\arcmin$ were observed in RI75I86 with a
2048$\times$2048 CCD. Landolt (1992) broad-band standards and Massey et al. 
(1988) spectrophotometric standards were observed in every bandpass
across all the airmass ranges observed for each night, and
Wolf 1346 was always observed in RI75I86 to calibrate the nonstandard
I-filters. The data were reduced and
calibrated in the usual manner with IRAF; aperture photometry was then
measured for the objects in those fields with published redshifts using 
APPHOT.  We obtained photometry with 0.1 magnitude error or better
in all six passbands for 42 of these galaxies.

\subsection{Field Galaxy Data}

Spectroscopic redshifts were also obtained for a number of galaxies 
in the HOGPW survey.  38 galaxies were measured using the Kitt Peak 
4-meter telescope, in parallel with observations of
quasar candidates in the survey fields (Hall et al. 1996b).  We refer
the reader to that work for the details of
data acquisition and reduction.  In summary,
a number of observational setups were used: a single longslit,
multislits, and the HYDRA multifiber positioner and bench spectrograph.
23 additional spectra were obtained with the Blue Channel
Spectrograph on the Multiple Mirror Telescope.
Again, the usual procedures in IRAF were used for data reduction and
extraction of spectra.  Redshifts were obtained by inspection of 
well known emission and absorption features (such as the Balmer lines,
[OII]$\lambda$3727\AA , [OIII]$\lambda\lambda$4959,5007\AA , and 
the 4000\AA\ break) or by
cross-correlation using the XCOR task in IRAF.  All 61 galaxies have
UBVRI75I86 photometry, which are calibrated with the procedures and
parameters described in HOGPW.

\subsection{Photometric Redshifts}

Identifying the photometric redshift and spectral type of any given
galaxy involves finding its best matching redshift-type template.
Clearly, the simplest definition of a match is that the template
SED matches the data more closely than any other template.
This is rarely a trivial criterion to meet, however, since every
real galaxy is at least slightly different from any template galaxy.
In each color, there is a difference between a template SED's color
and the observed color.  

One straightforward scheme to determine the closest match between
an observed galaxy and a template would be some kind of least-squares
method --- comparing the observed flux in each waveband with the
flux predicted by a template.  We tested several variations of
this method, but it proved not to be optimal for the purposes of our study.
One reason is that we want to identify the redshift and spectral type of
a galaxy without any knowledge of its apparent magnitude.  Thus we must 
use the comparison of galaxy colors, rather than passband fluxes.
Also, our tests showed that a least-squares technique 
can estimate the redshift of a galaxy reliably,
but too often misclassifies its spectral type.

A more successful approach, which we ultimately adopted, was to look
for matches by seeking the smallest maximum difference between the
observed and template colors.  (Such a method might loosely
be compared with a least-squares method the way a Kolmogorov-Smirnov
test might compare with a $\chi^2$-test, when measuring the quality
of a model fitted to data.)  This kind of strategy yielded the best
results in predicting both the redshift and spectral type of a 
galaxy, independent of its apparent magnitude.  Examination
of the color distributions of the model SEDs, combined with empirical
tests of choosing matches using various procedures, led us to
the following algorithm for selecting matches most accurately:

$<$i$>$ Each model-vs.-data comparison yields a set of
$\Delta_{i-j} \equiv (i-j)_{galaxy} - (i-j)_{model}$,
where $i$-$j=$ U-B, B-V, V-R, R-I75, I75-I86.  The most likely match is 
the comparison that yields the minimum $largest\ \Delta_{i-j}$.
The value of such a maximum for an accurately matched redshift-type
pair is typically $\Delta_{i-j}max \sim $0.1 to 0.2 magnitude.

$<$ii$>$ If $\Delta_{i-j}max$ is large (i.e. $>$0.2) for all possible
matches, the comparisons
that yield the minimum $second-largest\ \Delta_{i-j}$ are also reviewed.
This step takes into account the possibility that photometry for that
galaxy may have been anomalously affected in one filter, perhaps by an
emission line or other spectral feature.  Comparisons which yield very
large values of $\Delta_{i-j}max$ ($>$0.5), however, are not considered
possible matches.

$<$iii$>$ If two or more matches have similar $\Delta_{i-j}max$, the
best match is usually the one where the sum of $\Delta_{i-j}$'s is
closest to zero.  Matches with similar $\Delta_{i-j}$ characteristics are
almost always of the same galaxy spectral type, with slightly different
redshifts; in cases where it appears $\Sigma\ \Delta_{i-j}$ would be
closest to zero in an interpolation between two adjacent redshift steps,
the midpoint between those two steps is designated the most likely redshift.

$<$iv$>$ Finally, if two matches meet the above criteria with the same
accuracy, the one most closely matching in (U-B) or (I75-I86),
for objects with likely redshift less than or greater than 0.5 respectively,
is designated the more likely match.  This condition is based on the fact
that the separations in $F_{\lambda}$ of galaxy SEDs are widest in the 
UV and far-red wavelengths (see Figure 1).

\section{Discussion}

A computer program called ``GetZ'' was written
which automates the above analysis.
The colors of the 103 sample galaxies were then run through the program.
We present in Figure 4 a direct comparison of $z_p$, the photometrically
determined redshifts output by ``GetZ,'' 
with the spectroscopically measured redshifts $z_s$.
The diagonal line is not a fit to the data, but the locus of exactly
perfect correspondence (i.e. $z_s=z_p$).  The dispersion measure
$z_p-z_s$ for the entire sample is plotted in histogram form in Figure 5.

The rms deviation of all the galaxies in the sample, $\sigma_z$, is 0.053.
87\% of the galaxies have $\vert z_p-z_s \vert < $ 0.1.  As Figure 5
shows, the distribution of $z_p-z_s$ is essentially symmetric about 
$z_p-z_s = $ 0; in fact, $\Sigma(z_p-z_s) = $ 0.002, and the
error distribution is consistent with a Gaussian distribution.
These results are consistent with the accuracy we expected, and with those
in the literature.  

\subsection{The Effects of ``E+A'' Galaxies}

The cluster galaxy subsample had a slightly higher $\sigma_z$ than 
the HOGPW subsample (0.060 vs. 0.046).  Statistically, the lower 
accuracy of the cluster sample is barely significant,
but it is instructive to examine the cause of this slight discrepancy.

The cluster samples were all taken near the centers of rich clusters of
galaxies which exhibit the so-called Butcher-Oemler effect 
(Butcher \& Oemler 1978).  
Not surprisingly, a number of post-starburst, or ``E+A'' galaxies 
(Dressler \& Gunn 1983; Couch \& Sharples 1987;
Liu \& Green 1996), were observed in the CCD image fields which we used
to obtain photometry of the sample galaxies.
E+A galaxies have a spectrum characterized by strong Balmer absorption,
weak or no line emission, and the earmarks of an old stellar population.
Their spectral energy distributions are thus a hybrid, typically
with colors bluer than ellipticals but redder than late-type 
spirals.  $A\ priori$, then, it seems plausible that E+A galaxies could 
systematically confuse the redshift-spectral type comparison scheme,
since their colors are neither truly spiral nor truly elliptical.

To test this hypothesis, we have selected the E+A galaxies
in the cluster sample -- those objected which were 
designated ``a'' or ``A'' by Dressler \& Gunn (1992), or ``E+A'' by
Fabricant et al. (1991) -- and plotted their $z_p-z_s$
distributions separately (with shaded bars) on Figure 5.  The E+A's
are clearly less accurately identified than the other galaxy types.
$\sigma_z$ for the E+A's is 0.097 for 13 objects; and it is half that 
value (0.047 for 29 objects) for the rest of the cluster sample if the
E+A's are excluded.  Figure 5 shows also that the $z_p$'s
which are the worst underestimates of the true redshifts (i.e. the
four objects where $z_p-z_s < -$0.12) are all E+A galaxies.
Although about half (6/13) of the E+A's were successfully 
measured to within 0.05 of their spectroscopic redshifts, our results
demonstrate that E+A galaxies can increase the uncertainty of 
photometric redshift measurements and cause systematic underestimates
of $z_p$.  

Can this problem be overcome by simply creating one or more templates 
of E+A galaxies and adding them to the photometric redshift algorithm?
Unfortunately, this is not a solution.  As shown by Liu \& Green (1996),
E+A galaxies are a rather heterogeneous population which cannot be
characterized by one or even a few templates.  Their SEDs vary widely
depending on the strength of the decaying starburst, and on the underlying
old stellar population.  Furthermore, no UV spectrophotometry of E+A
galaxies exist in the literature; 
E+A templates can therefore be produced only
by stellar population models, which we have explicitly avoided in this
work.  Since E+A galaxies appear to be rare in the field, however
(Zabludoff et al. 1996), the random error contributed by E+A's is likely 
to be negligible in field galaxy studies with photometric redshifts.

\subsection{Identifications With Incomplete Data}

In any galaxy survey where a photometric redshift scheme such as
ours is likely to be applied, some portion of the galaxies will have
incomplete color information, such as a non-detection in one or more
filters.  This is especially likely for faint early-type
galaxies; as Table 1 shows, if an unevolved elliptical is 
detected in R at a given apparent magnitude, the survey data must extend 
at least two magnitudes fainter in U to be detected if the galaxy is at z=0,
and four magnitudes if it is at z=0.4.  It is important to know if redshift
identifications are still accurate or possible with missing information,
particularly U-band.

We can get some idea of what results to expect with color-color diagrams.
From Figure 1, it is clear that lack of U-band data does not significantly
affect the separation of high-redshift galaxies from each other; the 
problem with losing blue and UV data lies in confusing low-redshift, redder
galaxies from higher-redshift bluer galaxies.  If we exclude 
U-band data and use all the photometry from B redward (see Figure 6), 
the colors of E/S0 and Sab galaxies from 
0$\lesssim$z$\lesssim$0.25 are essentially
degenerate with those of Sbc galaxies from 0.1$\lesssim$z$\lesssim$0.4 and 
Scd galaxies from 0.2$\lesssim$z$\lesssim$0.6.  If both U and B data are
missing, the risk of low-redshift confusion is even greater.

Empirical tests confirm the problems with measuring 
photometric redshifts which are
suggested by the color-color plots in Figure 6.  The BVRI75I86 colors for the
sample galaxies were input into the ``GetZ'' program, and $z_p$ was
determined for each object as before.  This time, $\sigma_z = $ 0.101,
double the value of $\sigma_z$ computed with U-band data.  Furthermore,
galaxies which were identified as spiral or irregular with z$<$0.4 had
$\sigma_z = $ 0.160; the scatter is much larger, and there is a systematic
tendency to misidentify Scd and Irr galaxies 
near z$\sim$0.3 as Sbc galaxies near z$\sim$0.05.  

On the other hand, 
$\sigma_z$ for all galaxies with z$>$0.4 and for early-type galaxies with
z$>$0.25 were 0.053 and 0.057 respectively.  Apparently there is little
reduction in the typical accuracies of $z_p$ for those subsamples despite
the lack of U-band data; this is also predicted by the color-color plots,
and is the fortuitous result of the 4000\AA\ break being redshifted into
the V and R bands, away from the U-band.
If it were possible, then, to select galaxies which are definitely early-type
or at high redshift, our photometric method can still be effective for
determining redshifts, even without the U-band data which is so
critical for measuring later-type galaxies at lower redshifts.  

\section{Conclusions}

We have shown that our photometric redshift method based on the 
broad-band colors
UBVR, I75 and I86 can determine redshifts to a typical 
accuracy of z $=\pm$ 0.05 for field and cluster galaxies,
and approximate their spectral
 types as well.  This result extends the increasing amount of
literature that confirms the validity of using multicolor broad-band 
photometry to obtain a redshift distribution for samples of galaxies.
Since our system relies on colors alone, it is
somewhat more versatile than systems which are dependent on other
galaxy parameters such as apparent magnitude.  Thus, it can be (and has
been -- see Liu et al. 1998) applied to a wider range 
of astrophysical problems, such as field
galaxy evolution as a function of redshift.

It should be emphasized that any photometric redshift system is most
effectively used as a $statistical$ tool for measuring the redshift
distribution of a galaxy sample, rather than for assigning unambiguous
redshifts to individual galaxies.  
Just as a significant fraction of galaxies defy straightforward
classification on the morphological Hubble sequence, galaxies which are
not spectrophotometrically ``normal'' -- such as E+A galaxies -- can
cause systematic errors in redshift and spectral type determinations.
Incomplete data, such as a lack of U-band photometry, can also produce
serious mistakes in $z_p$ measurements; this can be 
overcome, however, by selecting samples of higher-redshift and/or 
early-type galaxies for study.  With a healthy awareness of the method's
strengths and weaknesses, and a careful attention to detail, using 
multicolors to obtain galaxy redshifts and spectral types
is a feasible and powerful technique for use in the study 
of galaxy populations.

\acknowledgments
  
We thank Pat Hall and Pat Osmer
for invaluable assistance in obtaining much of the
spectroscopic comparison data, and 
Andy Connolly and David Koo for helpful discussions.
C. L. gratefully acknowledges support from NASA grant NGT-50758.

\clearpage

\begin{deluxetable}{lccccc}
\scriptsize
\tablewidth{0pc}
\tablecaption{E/S0 template galaxy colors}
\tablehead{
\colhead{z}         & \colhead{U-B}     &
\colhead{B-V}       & \colhead{V-R}     &
\colhead{R-I75}     & \colhead{I75-I86}
}

\startdata

0.000 & 0.539 & 0.924 & 0.590 & 0.143 & 0.243 \nl  
0.025 & 0.490 & 1.016 & 0.604 & 0.150 & 0.229 \nl  
0.050 & 0.436 & 1.115 & 0.613 & 0.154 & 0.226 \nl  
0.075 & 0.393 & 1.216 & 0.627 & 0.161 & 0.228 \nl  
0.100 & 0.362 & 1.310 & 0.643 & 0.180 & 0.230 \nl  
0.125 & 0.342 & 1.389 & 0.668 & 0.184 & 0.260 \nl  
0.150 & 0.341 & 1.452 & 0.699 & 0.194 & 0.280 \nl  
0.175 & 0.360 & 1.503 & 0.728 & 0.221 & 0.278 \nl  
0.200 & 0.406 & 1.543 & 0.761 & 0.240 & 0.279 \nl  
0.225 & 0.480 & 1.568 & 0.806 & 0.250 & 0.284 \nl  
0.250 & 0.572 & 1.565 & 0.874 & 0.257 & 0.299 \nl  
0.275 & 0.668 & 1.539 & 0.964 & 0.257 & 0.315 \nl  
0.300 & 0.765 & 1.506 & 1.059 & 0.251 & 0.330 \nl  
0.325 & 0.867 & 1.478 & 1.145 & 0.256 & 0.335 \nl  
0.350 & 0.969 & 1.456 & 1.223 & 0.265 & 0.346 \nl  
0.375 & 1.058 & 1.439 & 1.292 & 0.270 & 0.373 \nl  
0.400 & 1.122 & 1.433 & 1.347 & 0.282 & 0.403 \nl  
0.425 & 1.155 & 1.441 & 1.387 & 0.311 & 0.423 \nl  
0.450 & 1.153 & 1.461 & 1.418 & 0.354 & 0.428 \nl  
0.475 & 1.114 & 1.495 & 1.441 & 0.398 & 0.428 \nl  
0.500 & 1.039 & 1.546 & 1.453 & 0.454 & 0.415 \nl  
0.525 & 0.935 & 1.611 & 1.444 & 0.526 & 0.404 \nl  
0.550 & 0.812 & 1.685 & 1.409 & 0.607 & 0.403 \nl  
0.575 & 0.680 & 1.760 & 1.364 & 0.685 & 0.408 \nl  
0.600 & 0.544 & 1.833 & 1.322 & 0.749 & 0.417 \nl  
0.625 & 0.412 & 1.902 & 1.291 & 0.786 & 0.439 \nl  
0.650 & 0.287 & 1.968 & 1.267 & 0.801 & 0.480 \nl  
0.675 & 0.174 & 2.025 & 1.247 & 0.810 & 0.531 \nl  
0.700 & 0.075 & 2.070 & 1.233 & 0.811 & 0.597 \nl  
0.725 &-0.009 & 2.101 & 1.228 & 0.809 & 0.673 \nl  
0.750 &-0.077 & 2.116 & 1.235 & 0.806 & 0.749 \nl  
0.775 &-0.131 & 2.115 & 1.253 & 0.798 & 0.820 \nl  
0.800 &-0.170 & 2.097 & 1.281 & 0.772 & 0.895 \nl  
0.825 &-0.194 & 2.060 & 1.321 & 0.726 & 0.975 \nl  
0.850 &-0.207 & 2.007 & 1.371 & 0.663 & 1.055 \nl  
0.875 &-0.209 & 1.936 & 1.432 & 0.599 & 1.122 \nl  
0.900 &-0.206 & 1.854 & 1.499 & 0.537 & 1.173 \nl  
0.925 &-0.200 & 1.763 & 1.565 & 0.478 & 1.218 \nl  
0.950 &-0.191 & 1.664 & 1.629 & 0.431 & 1.247 \nl  
0.975 &-0.184 & 1.563 & 1.689 & 0.418 & 1.243 \nl  
1.000 &-0.178 & 1.459 & 1.749 & 0.437 & 1.209 \nl  

\enddata
\end{deluxetable}

\begin{deluxetable}{lccccc}
\scriptsize
\tablewidth{0pc}
\tablecaption{Sab template galaxy colors}
\tablehead{
\colhead{z}         & \colhead{U-B}     &
\colhead{B-V}       & \colhead{V-R}     &
\colhead{R-I75}     & \colhead{I75-I86}
}

\startdata

0.000 & 0.364 & 0.853 & 0.595 & 0.173 & 0.252 \nl
0.025 & 0.385 & 0.903 & 0.617 & 0.173 & 0.260 \nl
0.050 & 0.379 & 0.966 & 0.634 & 0.169 & 0.285 \nl
0.075 & 0.354 & 1.039 & 0.646 & 0.177 & 0.303 \nl
0.100 & 0.316 & 1.114 & 0.654 & 0.205 & 0.295 \nl
0.125 & 0.275 & 1.188 & 0.662 & 0.215 & 0.307 \nl
0.150 & 0.241 & 1.259 & 0.671 & 0.230 & 0.309 \nl
0.175 & 0.218 & 1.328 & 0.676 & 0.257 & 0.297 \nl
0.200 & 0.206 & 1.390 & 0.684 & 0.269 & 0.299 \nl
0.225 & 0.207 & 1.446 & 0.696 & 0.289 & 0.296 \nl
0.250 & 0.226 & 1.480 & 0.725 & 0.306 & 0.309 \nl
0.275 & 0.257 & 1.496 & 0.769 & 0.310 & 0.333 \nl
0.300 & 0.287 & 1.500 & 0.823 & 0.300 & 0.354 \nl
0.325 & 0.312 & 1.496 & 0.882 & 0.295 & 0.366 \nl
0.350 & 0.333 & 1.479 & 0.946 & 0.285 & 0.387 \nl
0.375 & 0.343 & 1.453 & 1.011 & 0.273 & 0.413 \nl
0.400 & 0.338 & 1.425 & 1.074 & 0.271 & 0.439 \nl
0.425 & 0.327 & 1.398 & 1.134 & 0.277 & 0.460 \nl
0.450 & 0.316 & 1.374 & 1.192 & 0.289 & 0.474 \nl
0.475 & 0.302 & 1.356 & 1.244 & 0.302 & 0.481 \nl
0.500 & 0.279 & 1.346 & 1.291 & 0.326 & 0.472 \nl
0.525 & 0.240 & 1.343 & 1.324 & 0.368 & 0.457 \nl
0.550 & 0.188 & 1.351 & 1.335 & 0.420 & 0.440 \nl
0.575 & 0.129 & 1.371 & 1.327 & 0.472 & 0.426 \nl
0.600 & 0.067 & 1.393 & 1.311 & 0.524 & 0.413 \nl
0.625 & 0.001 & 1.415 & 1.293 & 0.566 & 0.412 \nl
0.650 &-0.065 & 1.433 & 1.271 & 0.602 & 0.425 \nl
0.675 &-0.126 & 1.448 & 1.240 & 0.643 & 0.439 \nl
0.700 &-0.181 & 1.457 & 1.205 & 0.679 & 0.465 \nl
0.725 &-0.232 & 1.454 & 1.176 & 0.716 & 0.497 \nl
0.750 &-0.279 & 1.440 & 1.157 & 0.751 & 0.534 \nl
0.775 &-0.321 & 1.418 & 1.146 & 0.782 & 0.570 \nl
0.800 &-0.355 & 1.393 & 1.137 & 0.799 & 0.616 \nl
0.825 &-0.382 & 1.366 & 1.132 & 0.796 & 0.672 \nl
0.850 &-0.404 & 1.337 & 1.128 & 0.776 & 0.738 \nl
0.875 &-0.419 & 1.297 & 1.134 & 0.749 & 0.799 \nl
0.900 &-0.429 & 1.252 & 1.150 & 0.721 & 0.848 \nl
0.925 &-0.433 & 1.205 & 1.177 & 0.668 & 0.906 \nl
0.950 &-0.431 & 1.155 & 1.209 & 0.596 & 0.974 \nl
0.975 &-0.428 & 1.107 & 1.237 & 0.548 & 1.018 \nl
1.000 &-0.427 & 1.062 & 1.255 & 0.510 & 1.057 \nl

\enddata
\end{deluxetable}

\begin{deluxetable}{lccccc}
\scriptsize
\tablewidth{0pc}
\tablecaption{Sbc template galaxy colors}
\tablehead{
\colhead{z}         & \colhead{U-B}     &
\colhead{B-V}       & \colhead{V-R}     &
\colhead{R-I75}     & \colhead{I75-I86}
}

\startdata

0.000 & 0.055 & 0.661 & 0.487 & 0.074 & 0.273 \nl
0.025 & 0.103 & 0.705 & 0.499 & 0.068 & 0.272 \nl
0.050 & 0.142 & 0.758 & 0.510 & 0.065 & 0.267 \nl
0.075 & 0.167 & 0.818 & 0.520 & 0.086 & 0.239 \nl
0.100 & 0.178 & 0.880 & 0.528 & 0.118 & 0.205 \nl
0.125 & 0.178 & 0.941 & 0.539 & 0.123 & 0.199 \nl
0.150 & 0.167 & 1.000 & 0.549 & 0.130 & 0.198 \nl
0.175 & 0.141 & 1.064 & 0.549 & 0.147 & 0.195 \nl
0.200 & 0.105 & 1.127 & 0.553 & 0.132 & 0.222 \nl
0.225 & 0.060 & 1.184 & 0.567 & 0.145 & 0.219 \nl
0.250 & 0.010 & 1.228 & 0.596 & 0.167 & 0.224 \nl
0.275 &-0.044 & 1.263 & 0.633 & 0.179 & 0.252 \nl
0.300 &-0.097 & 1.289 & 0.675 & 0.177 & 0.271 \nl
0.325 &-0.148 & 1.307 & 0.720 & 0.175 & 0.276 \nl
0.350 &-0.196 & 1.313 & 0.770 & 0.165 & 0.277 \nl
0.375 &-0.241 & 1.308 & 0.819 & 0.153 & 0.276 \nl
0.400 &-0.283 & 1.298 & 0.861 & 0.149 & 0.295 \nl
0.425 &-0.319 & 1.280 & 0.904 & 0.154 & 0.319 \nl
0.450 &-0.351 & 1.254 & 0.952 & 0.163 & 0.341 \nl
0.475 &-0.378 & 1.221 & 0.999 & 0.177 & 0.350 \nl
0.500 &-0.401 & 1.182 & 1.045 & 0.205 & 0.343 \nl
0.525 &-0.421 & 1.139 & 1.082 & 0.246 & 0.330 \nl
0.550 &-0.439 & 1.093 & 1.111 & 0.289 & 0.317 \nl
0.575 &-0.455 & 1.046 & 1.134 & 0.329 & 0.305 \nl
0.600 &-0.468 & 0.998 & 1.155 & 0.361 & 0.293 \nl
0.625 &-0.480 & 0.950 & 1.174 & 0.383 & 0.292 \nl
0.650 &-0.492 & 0.905 & 1.183 & 0.404 & 0.302 \nl
0.675 &-0.504 & 0.862 & 1.182 & 0.431 & 0.318 \nl
0.700 &-0.513 & 0.821 & 1.176 & 0.460 & 0.340 \nl
0.725 &-0.522 & 0.782 & 1.169 & 0.487 & 0.371 \nl
0.750 &-0.530 & 0.747 & 1.156 & 0.513 & 0.408 \nl
0.775 &-0.536 & 0.715 & 1.139 & 0.537 & 0.445 \nl
0.800 &-0.540 & 0.686 & 1.118 & 0.560 & 0.479 \nl
0.825 &-0.543 & 0.660 & 1.092 & 0.579 & 0.515 \nl
0.850 &-0.544 & 0.638 & 1.062 & 0.588 & 0.557 \nl
0.875 &-0.542 & 0.617 & 1.030 & 0.588 & 0.599 \nl
0.900 &-0.538 & 0.599 & 0.996 & 0.596 & 0.626 \nl
0.925 &-0.532 & 0.583 & 0.961 & 0.595 & 0.654 \nl
0.950 &-0.524 & 0.569 & 0.925 & 0.580 & 0.693 \nl
0.975 &-0.513 & 0.556 & 0.890 & 0.575 & 0.721 \nl
1.000 &-0.501 & 0.545 & 0.855 & 0.569 & 0.751 \nl

\enddata
\end{deluxetable}

\begin{deluxetable}{lccccc}
\scriptsize
\tablewidth{0pc}
\tablecaption{Scd template galaxy colors}
\tablehead{
\colhead{z}         & \colhead{U-B}     &
\colhead{B-V}       & \colhead{V-R}     &
\colhead{R-I75}     & \colhead{I75-I86}
}

\startdata

0.000 &-0.077 & 0.601 & 0.465 &-0.018 & 0.144 \nl
0.025 &-0.048 & 0.630 & 0.485 &-0.014 & 0.151 \nl
0.050 &-0.039 & 0.669 & 0.500 &-0.005 & 0.154 \nl
0.075 &-0.051 & 0.718 & 0.508 & 0.044 & 0.119 \nl
0.100 &-0.077 & 0.769 & 0.513 & 0.106 & 0.073 \nl
0.125 &-0.112 & 0.821 & 0.517 & 0.126 & 0.071 \nl
0.150 &-0.149 & 0.872 & 0.520 & 0.143 & 0.076 \nl
0.175 &-0.188 & 0.928 & 0.509 & 0.162 & 0.087 \nl
0.200 &-0.226 & 0.981 & 0.498 & 0.134 & 0.142 \nl
0.225 &-0.264 & 1.030 & 0.494 & 0.148 & 0.154 \nl
0.250 &-0.299 & 1.061 & 0.505 & 0.170 & 0.186 \nl
0.275 &-0.331 & 1.079 & 0.528 & 0.182 & 0.243 \nl
0.300 &-0.361 & 1.085 & 0.561 & 0.178 & 0.281 \nl
0.325 &-0.386 & 1.081 & 0.598 & 0.172 & 0.299 \nl
0.350 &-0.411 & 1.060 & 0.645 & 0.158 & 0.302 \nl
0.375 &-0.436 & 1.031 & 0.693 & 0.142 & 0.286 \nl
0.400 &-0.461 & 0.997 & 0.738 & 0.128 & 0.300 \nl
0.425 &-0.488 & 0.961 & 0.784 & 0.118 & 0.328 \nl
0.450 &-0.514 & 0.922 & 0.833 & 0.112 & 0.351 \nl
0.475 &-0.541 & 0.883 & 0.878 & 0.112 & 0.362 \nl
0.500 &-0.568 & 0.845 & 0.921 & 0.123 & 0.358 \nl
0.525 &-0.595 & 0.808 & 0.951 & 0.153 & 0.342 \nl
0.550 &-0.621 & 0.773 & 0.968 & 0.193 & 0.320 \nl
0.575 &-0.645 & 0.741 & 0.975 & 0.229 & 0.303 \nl
0.600 &-0.668 & 0.710 & 0.975 & 0.264 & 0.284 \nl
0.625 &-0.688 & 0.681 & 0.971 & 0.291 & 0.269 \nl
0.650 &-0.708 & 0.655 & 0.953 & 0.322 & 0.263 \nl
0.675 &-0.724 & 0.629 & 0.925 & 0.358 & 0.263 \nl
0.700 &-0.739 & 0.604 & 0.894 & 0.396 & 0.268 \nl
0.725 &-0.751 & 0.579 & 0.863 & 0.434 & 0.280 \nl
0.750 &-0.761 & 0.554 & 0.834 & 0.467 & 0.302 \nl
0.775 &-0.768 & 0.529 & 0.806 & 0.496 & 0.328 \nl
0.800 &-0.772 & 0.503 & 0.779 & 0.519 & 0.356 \nl
0.825 &-0.773 & 0.477 & 0.755 & 0.528 & 0.392 \nl
0.850 &-0.772 & 0.452 & 0.731 & 0.524 & 0.435 \nl
0.875 &-0.768 & 0.426 & 0.710 & 0.509 & 0.480 \nl
0.900 &-0.761 & 0.401 & 0.691 & 0.495 & 0.513 \nl
0.925 &-0.752 & 0.377 & 0.674 & 0.467 & 0.553 \nl
0.950 &-0.742 & 0.355 & 0.657 & 0.423 & 0.606 \nl
0.975 &-0.730 & 0.334 & 0.642 & 0.387 & 0.651 \nl
1.000 &-0.716 & 0.315 & 0.626 & 0.353 & 0.695 \nl

\enddata
\end{deluxetable}

\begin{deluxetable}{lccccc}
\scriptsize
\tablewidth{0pc}
\tablecaption{Irr template galaxy colors}
\tablehead{
\colhead{z}         & \colhead{U-B}     &
\colhead{B-V}       & \colhead{V-R}     &
\colhead{R-I75}     & \colhead{I75-I86}
}

\startdata

0.000 &-0.419 & 0.274 & 0.335 &-0.090 & 0.191 \nl
0.025 &-0.376 & 0.311 & 0.314 &-0.092 & 0.188 \nl
0.050 &-0.341 & 0.349 & 0.300 &-0.079 & 0.170 \nl
0.075 &-0.326 & 0.392 & 0.285 &-0.009 & 0.096 \nl
0.100 &-0.341 & 0.438 & 0.269 & 0.072 & 0.014 \nl
0.125 &-0.369 & 0.484 & 0.256 & 0.086 & 0.000 \nl
0.150 &-0.405 & 0.530 & 0.248 & 0.084 & 0.002 \nl
0.175 &-0.445 & 0.580 & 0.235 & 0.074 & 0.015 \nl
0.200 &-0.487 & 0.624 & 0.252 &-0.030 & 0.096 \nl
0.225 &-0.531 & 0.668 & 0.290 &-0.069 & 0.106 \nl
0.250 &-0.575 & 0.700 & 0.319 &-0.078 & 0.153 \nl
0.275 &-0.619 & 0.726 & 0.340 &-0.077 & 0.225 \nl
0.300 &-0.659 & 0.745 & 0.360 &-0.082 & 0.257 \nl
0.325 &-0.692 & 0.773 & 0.364 &-0.092 & 0.259 \nl
0.350 &-0.718 & 0.766 & 0.393 &-0.103 & 0.229 \nl
0.375 &-0.741 & 0.741 & 0.433 &-0.111 & 0.162 \nl
0.400 &-0.758 & 0.709 & 0.471 &-0.099 & 0.127 \nl
0.425 &-0.773 & 0.674 & 0.511 &-0.054 & 0.088 \nl
0.450 &-0.784 & 0.634 & 0.555 &-0.012 & 0.054 \nl
0.475 &-0.794 & 0.592 & 0.598 & 0.007 & 0.040 \nl
0.500 &-0.801 & 0.549 & 0.640 & 0.015 & 0.033 \nl
0.525 &-0.809 & 0.507 & 0.674 & 0.040 & 0.011 \nl
0.550 &-0.815 & 0.466 & 0.700 & 0.049 & 0.008 \nl
0.575 &-0.819 & 0.426 & 0.720 & 0.032 & 0.034 \nl
0.600 &-0.822 & 0.389 & 0.738 & 0.035 & 0.038 \nl
0.625 &-0.823 & 0.355 & 0.770 & 0.026 & 0.048 \nl
0.650 &-0.824 & 0.326 & 0.773 & 0.046 & 0.077 \nl
0.675 &-0.825 & 0.301 & 0.754 & 0.084 & 0.124 \nl
0.700 &-0.825 & 0.281 & 0.724 & 0.130 & 0.153 \nl
0.725 &-0.823 & 0.263 & 0.692 & 0.174 & 0.175 \nl
0.750 &-0.822 & 0.249 & 0.660 & 0.213 & 0.191 \nl
0.775 &-0.819 & 0.236 & 0.627 & 0.249 & 0.191 \nl
0.800 &-0.816 & 0.226 & 0.593 & 0.277 & 0.175 \nl
0.825 &-0.812 & 0.217 & 0.560 & 0.300 & 0.171 \nl
0.850 &-0.807 & 0.209 & 0.528 & 0.311 & 0.185 \nl
0.875 &-0.802 & 0.203 & 0.497 & 0.321 & 0.203 \nl
0.900 &-0.796 & 0.198 & 0.468 & 0.344 & 0.207 \nl
0.925 &-0.790 & 0.194 & 0.443 & 0.353 & 0.219 \nl
0.950 &-0.782 & 0.190 & 0.418 & 0.338 & 0.254 \nl
0.975 &-0.775 & 0.188 & 0.396 & 0.310 & 0.302 \nl
1.000 &-0.768 & 0.188 & 0.375 & 0.271 & 0.358 \nl

\enddata
\end{deluxetable}


{\bf Fig. 1 } Filter transmissions and CCD quantum efficiency vs. wavelength.

\rotate{\rotate{\rotate{\psfig{file=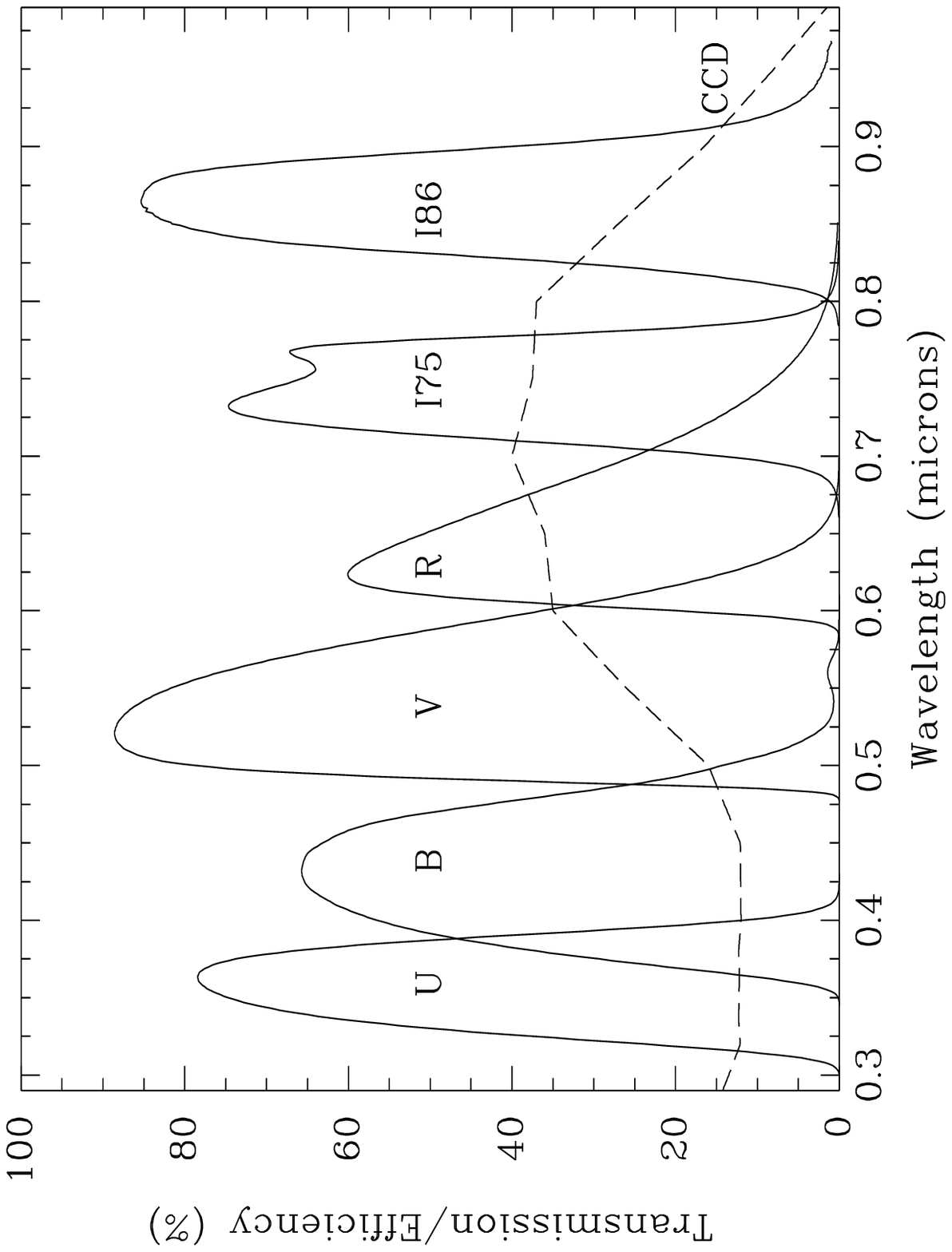,height=7in}}}}

\clearpage

{\bf Fig. 2 } Spectral energy distributions of the template galaxies,
in units of F$_{\lambda}$ normalized to unity at 5500\AA .

\plotone{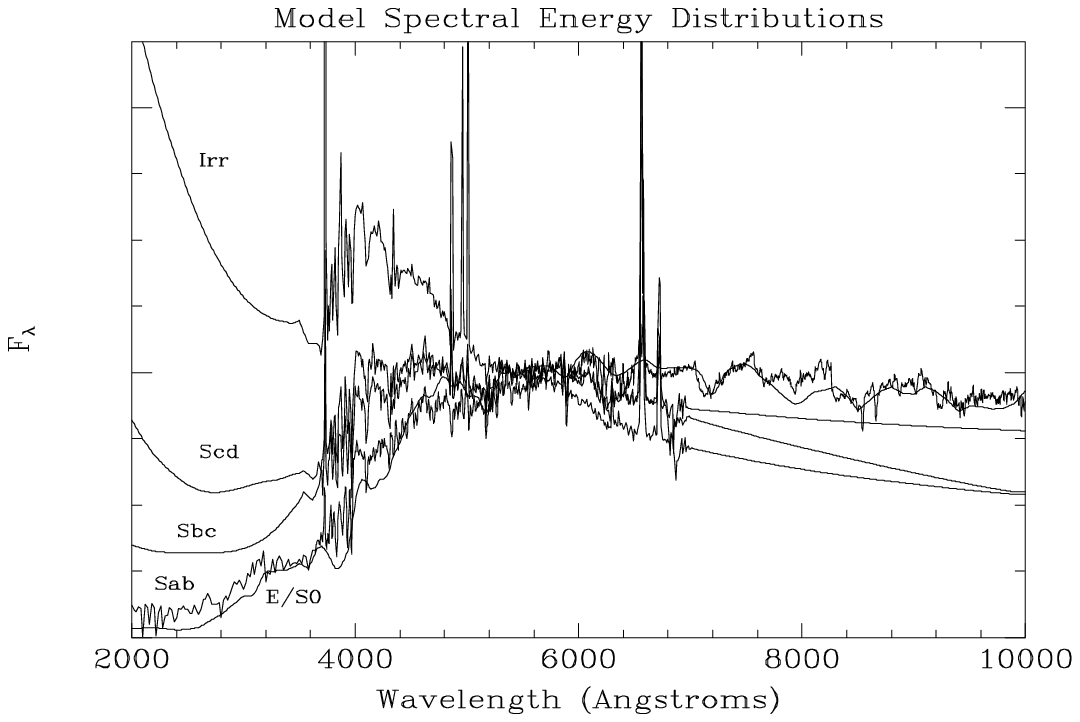}

\clearpage

{\bf Fig. 3 } Color evolutionary tracks for the template galaxy spectral types.
The tracks assume no luminosity evolution with redshift.  Each point on the 
tracks represents a stepwise increase in $z$ of 0.05.

\plotone{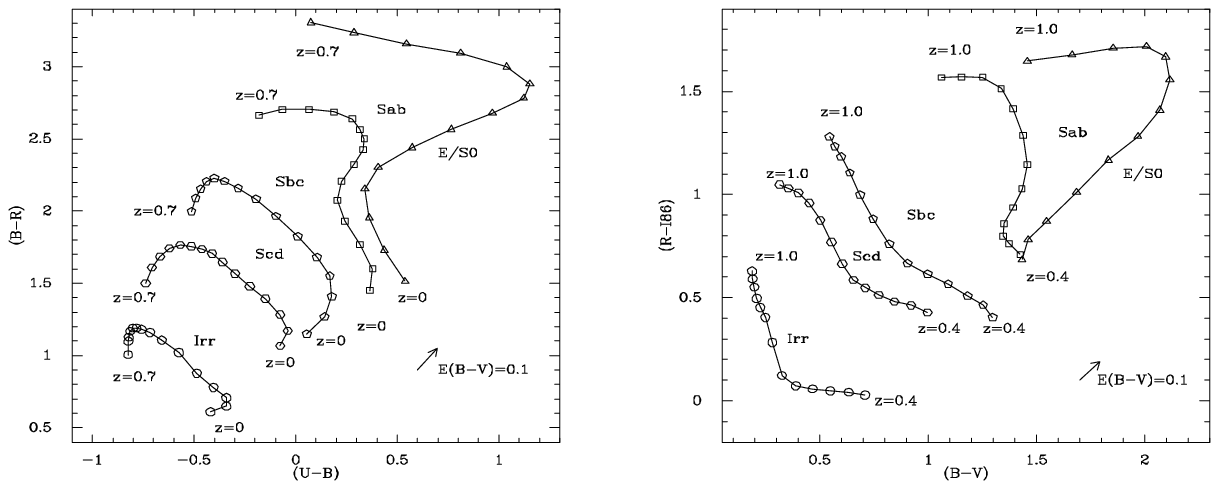}

\clearpage

{\bf Fig. 4 } Estimated photometric redshift vs. spectroscopically measured
redshift for galaxies from high-redshift clusters (``$+$'') and galaxies in
the HOGPW survey (``$\times$'').  The diagonal represents the locus of 
perfect agreement between the two different measurements.

\psfig{file=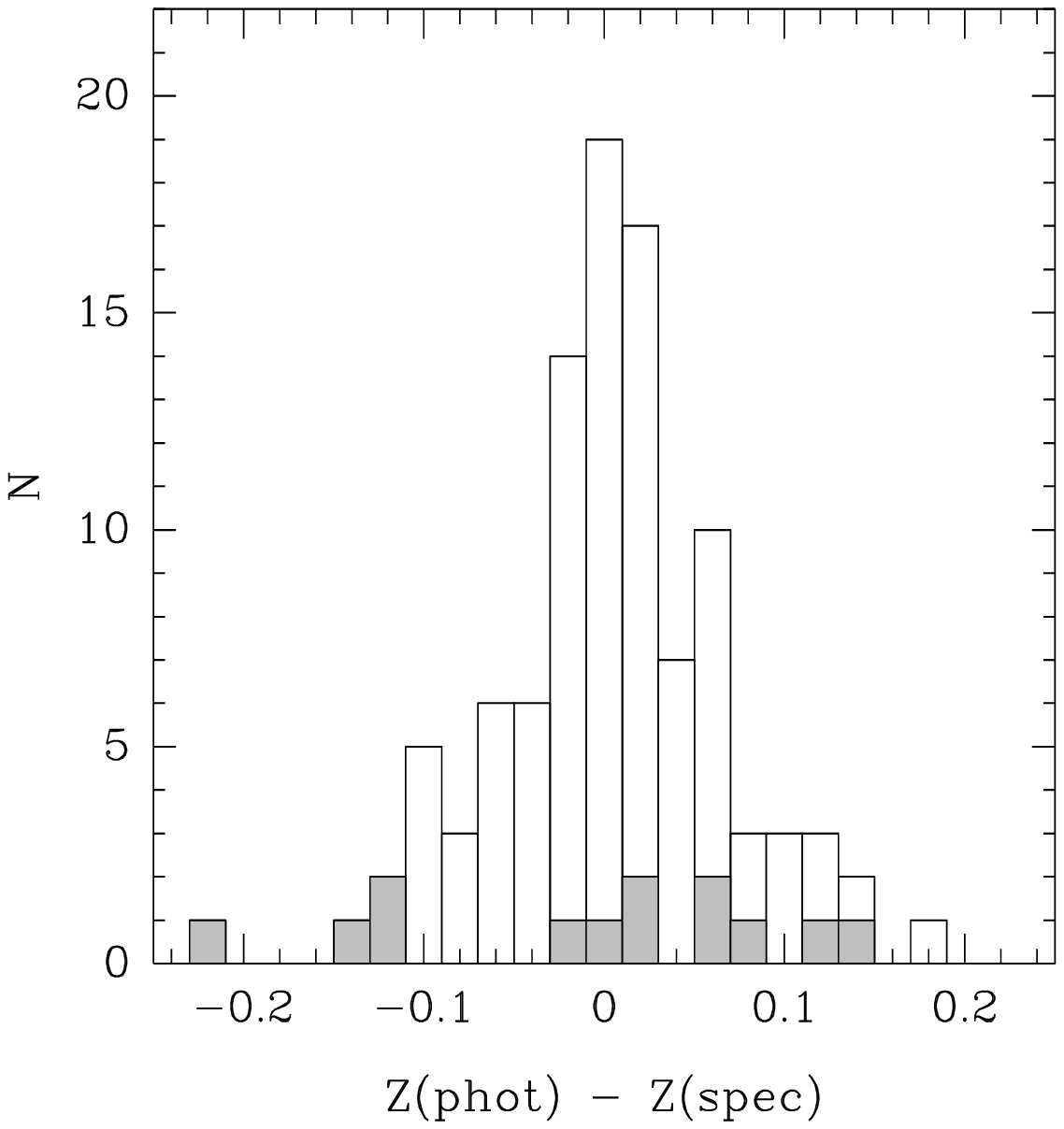,height=7.5in}

\clearpage

{\bf Fig. 5 } Distribution of photometric redshift errors.  
{\it Shaded bars:}
E+A galaxies in the sample, identified as described in the text.

\psfig{file=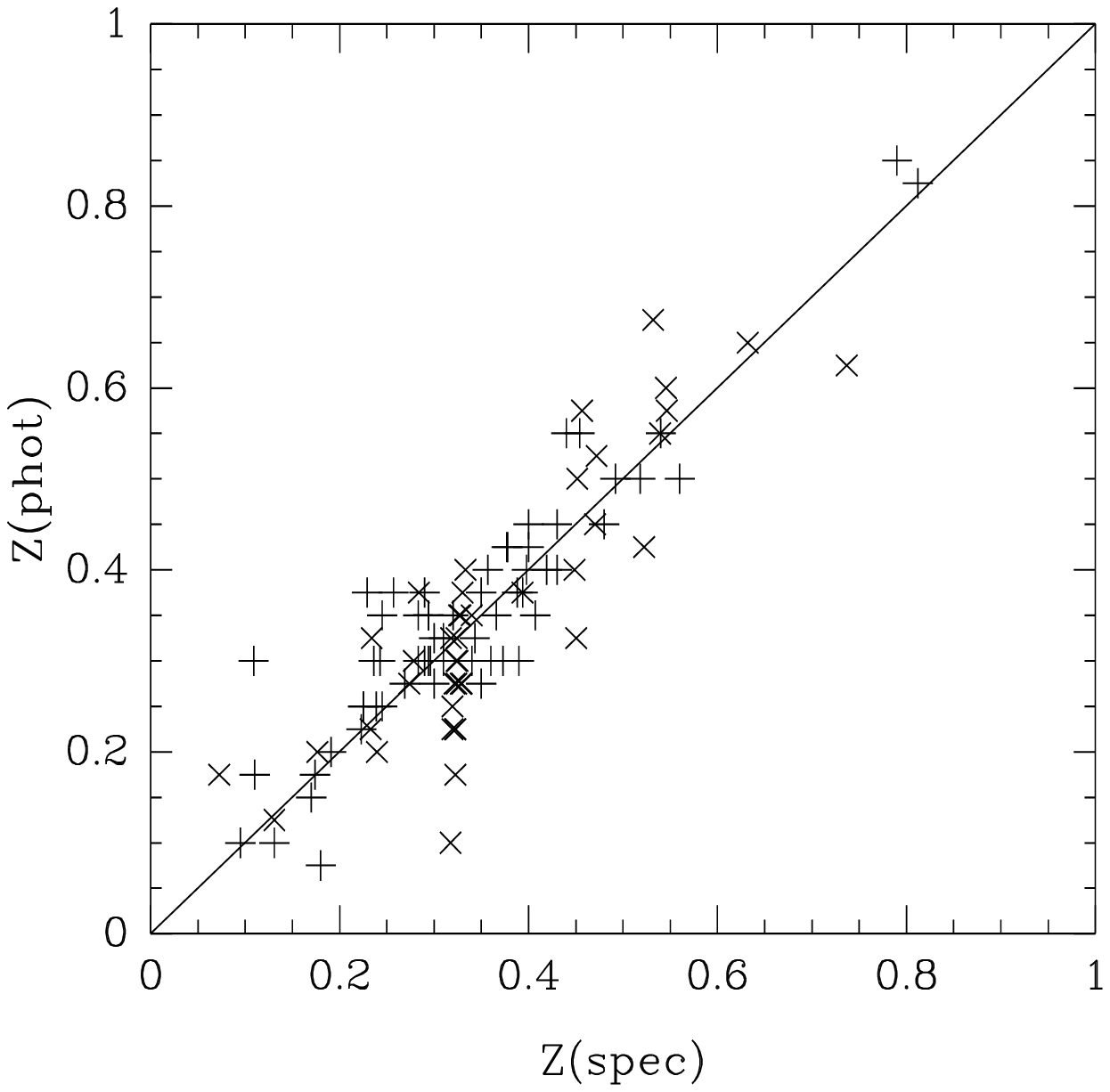,height=8in}

\clearpage

{\bf Fig. 6} Same as Fig. 3, but with color-color slices without U $(left)$
or U and B $(right)$ data.  Although degeneracies at low redshift are serious,
high redshift galaxy types are still well separated.

\plotone{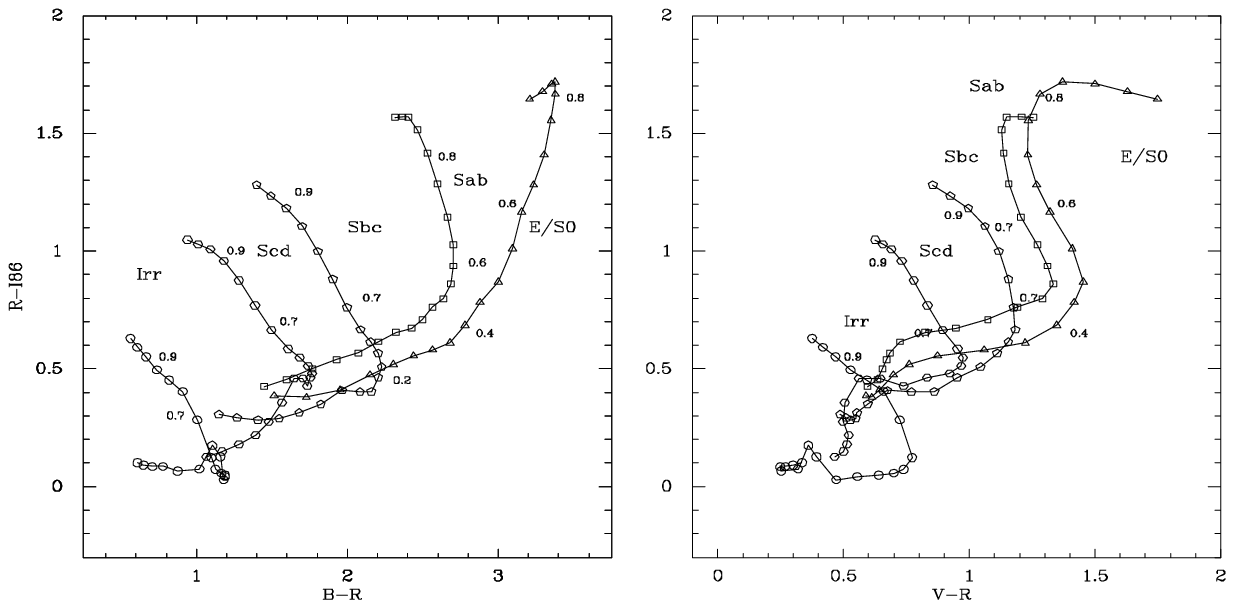}

\clearpage


\end{document}